\documentclass[cameraready]{Interspeech}

\title{Towards Event-Robust Acoustic Scene Classification}

\author[affiliation={1}, orcid=0009-0004-7062-7850, equalcontribution]{Yiqiang}{Cai}
\author[affiliation={1}, equalcontribution]{Bohan}{Hu}
\author[affiliation={2}]{Yu}{Yang}
\author[affiliation={3}]{Pengwei}{Lu}
\author[affiliation={1}, correspondingauthor]{Shengchen}{Li}
\author[affiliation={4}]{Xi}{Shao}

\address{
    $^1$ Xi'an Jiaotong-Liverpool University, Suzhou, China \\
    $^2$ Zhongdian Zhiheng Information Technology Service Co., Ltd, Nanjing, China \\
    $^3$ China Telecom Jiangsu Branch, Nanjing, China \\
    $^4$ Nanjing University of Posts and Telecommunications, Nanjing, China
}

\email{caiyiqaing0902@gmail.com, shengchen.li@xjtlu.edu.cn}


\usepackage{comment}

\begin{document}

\maketitle

\begin{abstract}
This paper introduces the Event-Shifted Acoustic Scene (ESAS) dataset, a novel benchmark for evaluating the robustness of Acoustic Scene Classification (ASC) systems against unknown sound events. Existing ASC datasets typically contain recordings of clean and consistent audio, while real-world environments often include diverse and unexpected sound events. To bridge this gap, ESAS simulates real-world acoustic variability by injecting foreground sound events into background scenes with the assistance of large language models. In this work, we present the construction methodology, dataset statistics, and evaluation protocols. Furthermore, a comprehensive evaluation of state-of-the-art ASC systems is conducted using the ESAS benchmark. Experimental results reveal that existing ASC models suffer significant performance degradation when facing the event-shift challenge. The introduction of the ESAS dataset aims to drive future research toward event-robust ASC.
\end{abstract}

\section{Introduction}
Acoustic Scene Classification (ASC) is a fundamental task in the field of computational sound scene analysis \cite{barchiesi2015acoustic, virtanen2018computational}. ASC aims to recognize the environment in which an audio recording was captured, such as a park, airport, or metro station. While ASC models have achieved remarkable progress with the help of large-scale datasets \cite{Mesaros2018, jeong2022cochlscene, bai2024description} and deep learning architectures \cite{koutini2021receptive, cai2024tf, Schmid2023}, real-world acoustic scenes are often far more complex than those represented in current benchmarks.

An acoustic scene typically consists of foreground sound events and background noise \cite{devalraju2022multiview}. In practical environments, the foreground sound events within a scene can vary drastically depending on time, season, and location \cite{bai2025apsipa}. For example, a park during the day may be dominated by children playing and birds chirping, whereas at night it may contain footsteps or traffic noise from nearby roads. Similarly, the same residential area may sound very different across regions or countries, reflecting distinct cultural or environmental activities. These variations cause what we refer to as \textit{event shift}—a phenomenon where the foreground events within an acoustic scene change significantly while the underlying scene category remains the same. Event shift is common and inevitable in the real world, and it poses a major challenge to ASC systems that often rely heavily on event-related information \cite{bear19_interspeech, tonami2021joint, hou2023cooperative}. Studying event shift is therefore crucial not only for improving robustness but also for revealing how acoustic scene features encode and separate background ambience from foreground sound events.

Previous research has explored other dimensions of acoustic variations, such as cross-city setting \cite{bai2024description, tan2024acoustic}, cross-device setting \cite{Heittola2020, zhang2025ddsc}, and time-variant setting \cite{bai2025apsipa}. These studies have provided valuable insights into geographic, channel and temporal mismatches. However, such factors are relatively limited to recording conditions or hardware settings. In contrast, event shift is a more general and fundamental problem, as it naturally occurs in all real-world acoustic environments regardless of device, city, or recording setup. It reflects the intrinsic dynamic nature of soundscapes, where the acoustic composition of a scene continually evolves with human activity and environmental context. Despite its importance, there is still no dedicated benchmark to evaluate the event robustness of ASC systems—that is, their ability to recognize scenes accurately even when unexpected or unfamiliar sound events occur.

Collecting large-scale acoustic scene recordings containing truly unknown events is extremely challenging. Real-world data acquisition requires extensive manual effort, and annotating transient or overlapping events is time-consuming and error-prone. Moreover, it is difficult to ensure a balanced distribution of event types and signal-to-noise ratios across different scenes. In this context, synthetic data offers a feasible alternative. For example, the DCASE dataset includes synthetic recordings for unseen devices to simulate channel variations \cite{Heittola2020}, and previous study has generated synthetic acoustic scenes for the joint task of ASC and Sound Event Detection (SED) \cite{bear19_interspeech}. However, these datasets either lack event-shift configurations or contain only a small number of samples, limiting their use for event-robust ASC evaluation.

To address these limitations, we introduce the Event-Shifted Acoustic Scene (ESAS) dataset, a benchmark specifically designed to study and evaluate the robustness of ASC systems under event-shift conditions. ESAS is constructed by mixing background scenes from CochlScene \cite{jeong2022cochlscene} with foreground sound events from FSD50K \cite{fonseca2021fsd50k}, producing polyphonic sound scenes with multiple overlapping events that reflect real-world acoustic complexity. To ensure semantic consistency between events and scenes, event-scene grouping is guided by a large language model (LLM). The dataset contains 13 scene classes and 96 event classes, preserving the original training, validation, and test splits in CochlScene. The training and validation sets include only background-only and known-event recordings, while unknown-event mixtures appear exclusively in the test set. In experiments, we conduct a comprehensive evaluation of several state-of-the-art (SOTA) ASC systems using the ESAS dataset. The experimental results reveal a critical vulnerability in current ASC methodologies: existing systems experience a severe collapse in classification accuracy when processing acoustic samples containing unknown sound events. These findings underscore the limitations of current representation learning paradigms and highlight the urgent necessity for developing fundamentally event-robust architectures.

The primary contributions of this paper are threefold. In this work, we:
\begin{itemize}
    \item propose a novel benchmark designed to simulate real-world event-shift conditions using LLM-guided semantic grouping.
    \item provide an empirical evaluation of multiple SOTA ASC systems under controlled event-shift conditions.
    \item expose a critical flaw in existing ASC models in the presence of unknown sound events, thereby defining a problem for future research in event-robust ASC.
\end{itemize}

\begin{figure}
    \centering
    \includegraphics[width=\linewidth]{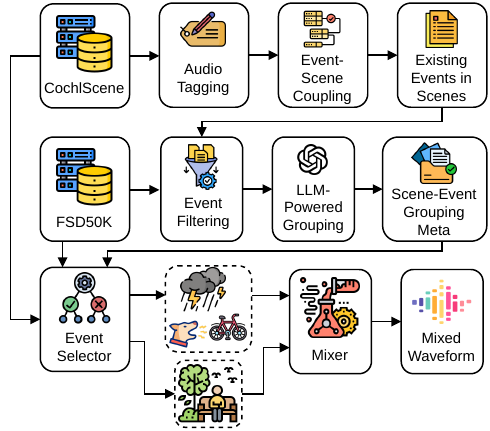}
    \caption{Overview of the ESAS dataset construction pipeline.}
    \label{fig:mix_pipeline}
\end{figure}

\section{Event-Shifted Acoustic Scene Dataset}
The Event-Shifted Acoustic Scene (ESAS) dataset is developed to evaluate the robustness of ASC systems under the presence of unknown sound events. It combines background recordings from CochlScene with foreground events from FSD50K. With the help of Large Language Model (LLM), we produce polyphonic mixtures that simulate natural soundscapes with controllable levels of event interference. The dataset is available at \url{https://doi.org/10.5281/zenodo.21317541}, and the code is available at \url{https://github.com/yqcai888/Interspeech2026_ESAS}.

\subsection{Dataset Construction}
Clips from CochlScene \cite{jeong2022cochlscene} are treated as background recordings. CochlScene is a large-scale acoustic scene dataset collected by using a crowdsourcing platform. It contains 13 scene classes captured from various urban and indoor locations. Each audio clip is 10 seconds long, recorded at 44.1 kHz, and provided with official train, validation, and test splits. Sound event samples are sourced from FSD50K \cite{fonseca2021fsd50k}. FSD50K is a human-labeled dataset of general sound events drawn from Freesound, containing approximately 51,000 audio clips across 200 sound event categories. The clips vary in duration (0.3-30 seconds) and cover a broad range of daily sounds. Its large coverage and clean labeling make it well suited for use as an event pool to synthesize event-rich soundscapes.

Figure \ref{fig:mix_pipeline} illustrates the synthesis pipeline for mixed acoustic scene recordings, designed to ensure the acoustic quality and semantic relevance of the injected foreground events. The process begins by analyzing the CochlScene recordings to detect any pre-existing sound events within the source audio, preventing the introduction of redundant foreground events into the original acoustic environment. To achieve this, we use the pre-trained BEATs model \cite{pmlr-v202-chen23ag}, a widely adopted architecture for robust audio tagging. Concurrently, the raw FSD50K dataset undergoes an event filtering process to isolate distinct sound events suitable for mixing by discarding recordings with low signal-to-noise ratios, ambiguous labels, or continuous background noises. These candidate events are divided into ``Known Events" (used in training and validation) and ``Unknown Events" (strictly isolated in the test set), which enables a two-stage failure analysis by decoupling the performance degradation caused by acoustic mixing effects (evaluated via known events) from the true vulnerability to out-of-distribution event shifts (isolated via unknown events). Subsequently, the identified scene contexts and candidate events are processed by an LLM-powered (GPT-4) scene-event grouping module, outputting structured metadata saved as a JSON file. It is worth noting that the LLM serves only as a constrained semantic filter to guarantee realistic acoustic combinations without affecting downstream evaluation labels. Finally, an event selector samples designated foreground events from the metadata, and a waveform mixer superimposes them onto the target background to yield the polyphonic waveform.

\begin{figure}
    \centering
    \includegraphics[width=\linewidth]{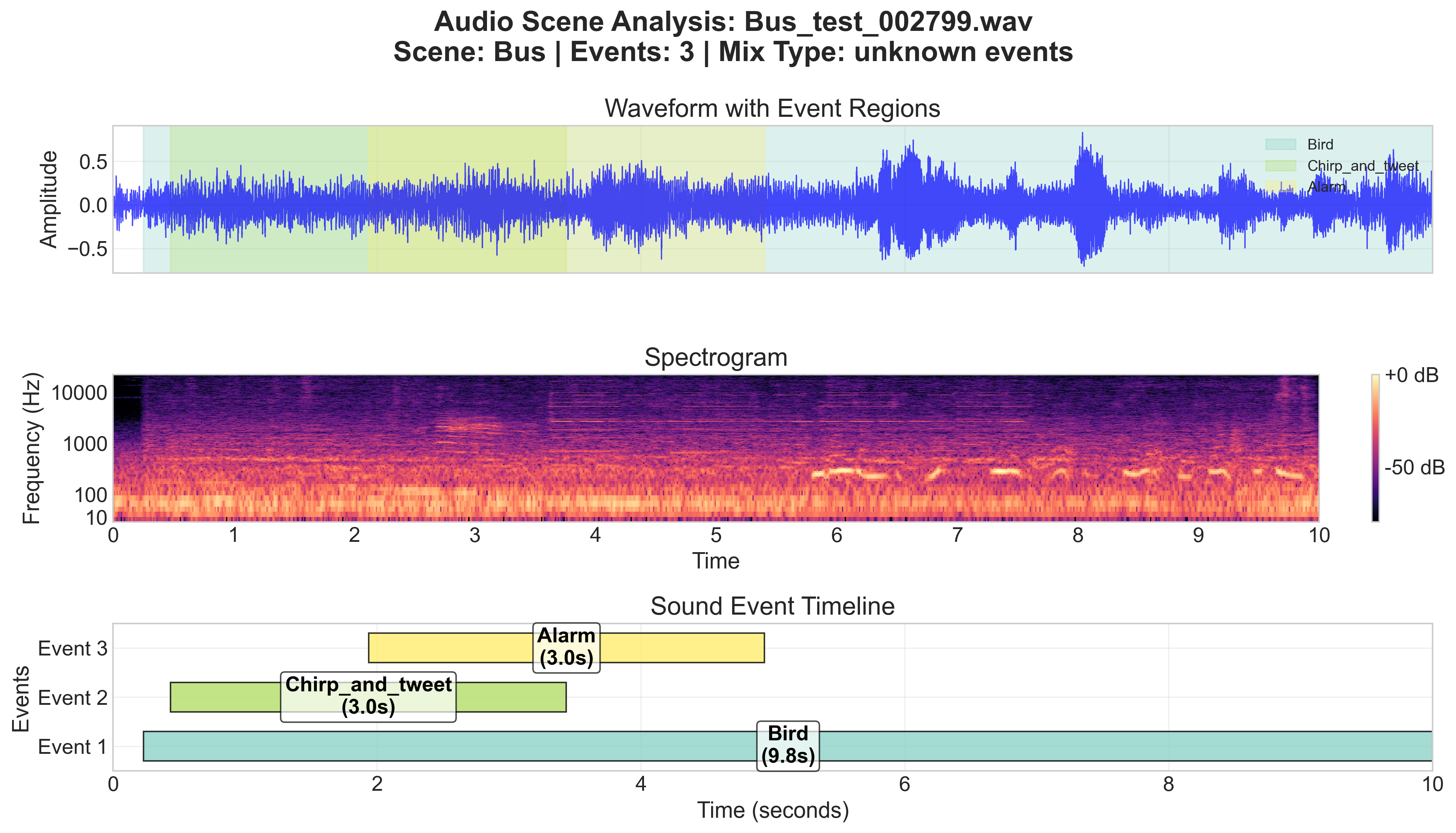}
    \caption{Illustration of a synthetic sample. Top: Waveform. Middle: Spectrogram. Bottom: Events being mixed.}
    \label{fig:mix_sample}
\end{figure}

For the configuration of the waveform mixer, we build upon the foundational mixing protocol established in \cite{bear19_interspeech}. Each synthesized sample is generated as a 10-second mono clip at a 44.1 kHz sampling rate. The number of distinct event classes per clip is uniformly sampled between 1 and 10, and their temporal positions are randomized such that the events are normally distributed throughout the clip. Furthermore, to further enhance acoustic variability, the foreground event audio undergoes data augmentation via time-stretching within a range of [0.8, 1.15] and pitch shifting between [-3, 3]. The final mixtures are blended at a randomized scene-to-event signal-to-noise ratio (SNR) ranging from -15dB to +15dB per clip. This protocol results in realistic polyphonic mixtures that accurately reflect the complex, overlapping nature of real-world acoustic environments. Finally, comprehensive generation metadata—including original scene identities, event labels and counts, precise timestamps, applied SNRs, dataset splits, and mix types—are preserved in a JSON metadata file for reproducibility.

Figure \ref{fig:mix_sample} illustrates a 10-second synthesized test sample, visualizing the injection of three unknown foreground events into a ``Bus" background acoustic scene. The waveform and spectrogram visualize the final acoustic result. The sound event timeline details the specific mixing configuration, showing overlapping ``Bird," ``Chirp and tweet," and ``Alarm" events of varying durations spanning the clip.

\subsection{Dataset Statistics}
Table \ref{tab:datasets} compares the proposed ESAS dataset with existing ASC benchmarks. ESAS comprises 211 hours of audio across 13 acoustic scene classes, maintaining the extensive scale of the foundational CochlScene dataset \cite{jeong2022cochlscene}. Unlike previous benchmarks that primarily target domain shifts related to recording locations \cite{Mesaros2018}, recording devices \cite{Heittola2020}, or time variation \cite{bai2024description}, ESAS uses a hybrid of real and synthetic audio specifically designed to evaluate the event shift problem. Each acoustic scene is associated with a unique set of background-only, known-event and unknown-event classes. Across all scenes, there are 27 known event classes and 69 unknown event classes, drawn from the FSD50K event pool.

Table \ref{tab:statistics} details the sample distribution across the ESAS dataset, which inherits the original training, validation, and test splits from CochlScene. The training and validation splits consist solely of background-only and known-event mixtures, while the test set includes additional unknown-event mixtures to simulate unexpected, real-world acoustic variability. On average, each mixture contains 4.38 events per clip, with an average event duration of 3.78 seconds. The test set maintains an approximate 1:1:1 ratio among background-only, known-event, and unknown-event samples for a balanced evaluation.

\begin{table}[]
    \centering
    \caption{Comparing ESAS to existing ASC Datasets.}
    \begin{tabular}{l|c c c c}
        \toprule
        \textbf{Dataset}& \textbf{Type}& \textbf{Duration}& \textbf{Classes}& \textbf{Shift} \\
        \midrule
        TAU19 \cite{Mesaros2018} & real& 40h& 10& city \\
        JSSED \cite{bear19_interspeech} & syn& 25h& 10& -\\
        TAU20 \cite{Heittola2020} & real\&syn& 64h& 10& device \\
        CAS23 \cite{bai2024description} & real& 24h& 10& city, time\\
        CochlScene \cite{jeong2022cochlscene} & real& 211h& 13& -\\
        ESAS (Ours) & real\&syn& 211h& 13& event\\
        \bottomrule
    \end{tabular}
    \label{tab:datasets}
\end{table}

\begin{table}[]
    \centering
    \caption{Distribution of audio clips across the ESAS dataset splits by mix type.}
    \begin{tabular}{c|ccc|c}
        \toprule
        \textbf{Mix Type} & \textbf{Train} & \textbf{Val} & \textbf{Test} & \textbf{Total} \\
        \midrule
        Background Only & 54,799 & 3,856 & 2,623 & 61,278 \\
        Known Events & 6,056 & 3,716 & 2,499 & 12,271 \\
        Unknown Events & - & - & 2,532 & 2,532 \\
        \midrule
        \textbf{Total} & 60,855 & 7,572 & 7,654 & 76,081 \\
        \bottomrule
    \end{tabular}
    \label{tab:statistics}
\end{table}

\subsection{Evaluation Protocol}
Following ASC conventions, the overall classification accuracy is used as the primary evaluation metric. To further highlight robustness, ESAS uses a three-tier evaluation protocol to decouple the sources of performance degradation. The background-only subset serves as a clean baseline for classifying undisturbed soundscapes. The known-event subset is then used to quantify the performance drop incurred by acoustic mixing and polyphony. Finally, testing on the unknown-event subset isolates the model's vulnerability to out-of-distribution event shifts. Comparing performance across these three conditions exposes the two-stage failure mechanism of ASC models, revealing the extent to which models are affected by event shift and quantifying their robustness to acoustic variability.

\section{Experiment}

\begin{table}[t]
    \centering
    \caption{Classification accuracy (\%) of baseline ASC systems (4 CNNs and 2 Transformers) evaluated on the ESAS dataset under background-only, known-event, and unknown-event conditions. * indicates models pre-trained on external datasets.}
    \begin{tabular}{l|c c c|c}
         \toprule
         \textbf{System}& \textbf{Background}& \textbf{Known}& \textbf{Unknown}& \textbf{Overall} \\
         \midrule
         TF-SepNet & 79.73 & 65.03 & 57.67 & 67.64 \\
         BC-ResNet & 78.28 & 66.35 & 59.18 & 68.06 \\
         GRU-CNN & 79.75 & 71.85 & 65.54 & 72.47 \\
         CP-Mobile & 79.53 & 73.38 & 68.10 & 73.74 \\
         \midrule
         BEATs$^{*}$ & 82.84 & 80.11 & 75.39 & 79.48 \\
         PaSST$^{*}$ & 84.27 & 79.93 & 75.19 & 79.85 \\
         \bottomrule
    \end{tabular}
    \label{tab:result}
\end{table}

\subsection{Baselines}
To comprehensively evaluate the vulnerability of current methodologies under event-shift conditions, we benchmark the ESAS dataset against a diverse set of state-of-the-art ASC models. Recognizing the dual necessity for both highly efficient and robust acoustic representation learning in real-world deployments, we selected baselines that span a wide spectrum of architectural paradigms—from lightweight convolutional neural networks (CNNs) to large-scale pre-trained Transformers. The evaluated systems are as follows:

\begin{itemize}
    \item TF-SepNet \cite{cai2024tf}: An efficient CNN architecture that uses 1D kernel designs to decouple temporal and frequency feature extraction.
    
    \item BC-ResNet \cite{kim21l_interspeech}: An architecture integrating broadcasted residual learning. Initially designed for highly efficient keyword spotting, it has proven to be a highly robust feature extractor for ASC.

    \item GRU-CNN \cite{Tan2025}: A low-complexity ASC model that combines a shallow CNN with a gated recurrent unit (GRU), configured to learn patterns across the frequency axis.
    
    \item CP-Mobile \cite{Schmid2023}: A highly compact architecture developed via a systematically controlled receptive field. It currently serves as the official baseline model for Task 1 of the recent DCASE Challenges.

    \item BEATs \cite{pmlr-v202-chen23ag}: A self-supervised audio pre-training framework that uses acoustic tokenizers. As a large-scale foundation model, it provides highly generalized audio representations and has demonstrated significant efficacy in ASC tasks \cite{Cai2024workshop}.
    
    \item PaSST \cite{koutini22_interspeech}: An adaptation of the Vision Transformer (ViT) architecture specifically designed for audio. Using a patchout mechanism during training, it achieved robust performance in complex acoustic scenes \cite{schmid2022knowledge}.
\end{itemize}

\subsection{Experimental Results}

\textbf{Performance of Different Baselines.} 
Table \ref{tab:result} details the classification accuracy of the evaluated state-of-the-art ASC systems across the three designated acoustic conditions: background-only, known-event, and unknown-event mixtures. As anticipated, all models establish a strong baseline performance on the clean, background-only scenes, achieving accuracies between 78.28\% and 84.27\%. However, the introduction of foreground sound events causes a consistent accuracy decline. Notably, performance degrades significantly even under the known-event condition. TF-SepNet experiences a substantial drop of 14.7\%, indicating that the presence of crowded, polyphonic acoustic mixtures can disrupts learned representations. Besides, the fundamental vulnerability of current representation learning paradigms is critically exposed under the unknown-event condition. Across the lightweight CNN architectures, accuracy suffers a severe collapse, dropping by up to 22 percentage points compared to their background-only baselines. Furthermore, while large-scale, externally pre-trained Transformer models (BEATs and PaSST) dictate the upper bound of overall performance and display greater relative resilience to known events, they still experience a clear, systematic degradation of roughly 7\% to 9\% percentage points when confronted with unseen sounds. Ultimately, these empirical results demonstrate that existing ASC systems struggle with polyphonic event interference, and broadly fail to generalize to the unpredictable event-shifted environments.

\begin{figure}
    \centering
    \includegraphics[width=\linewidth]{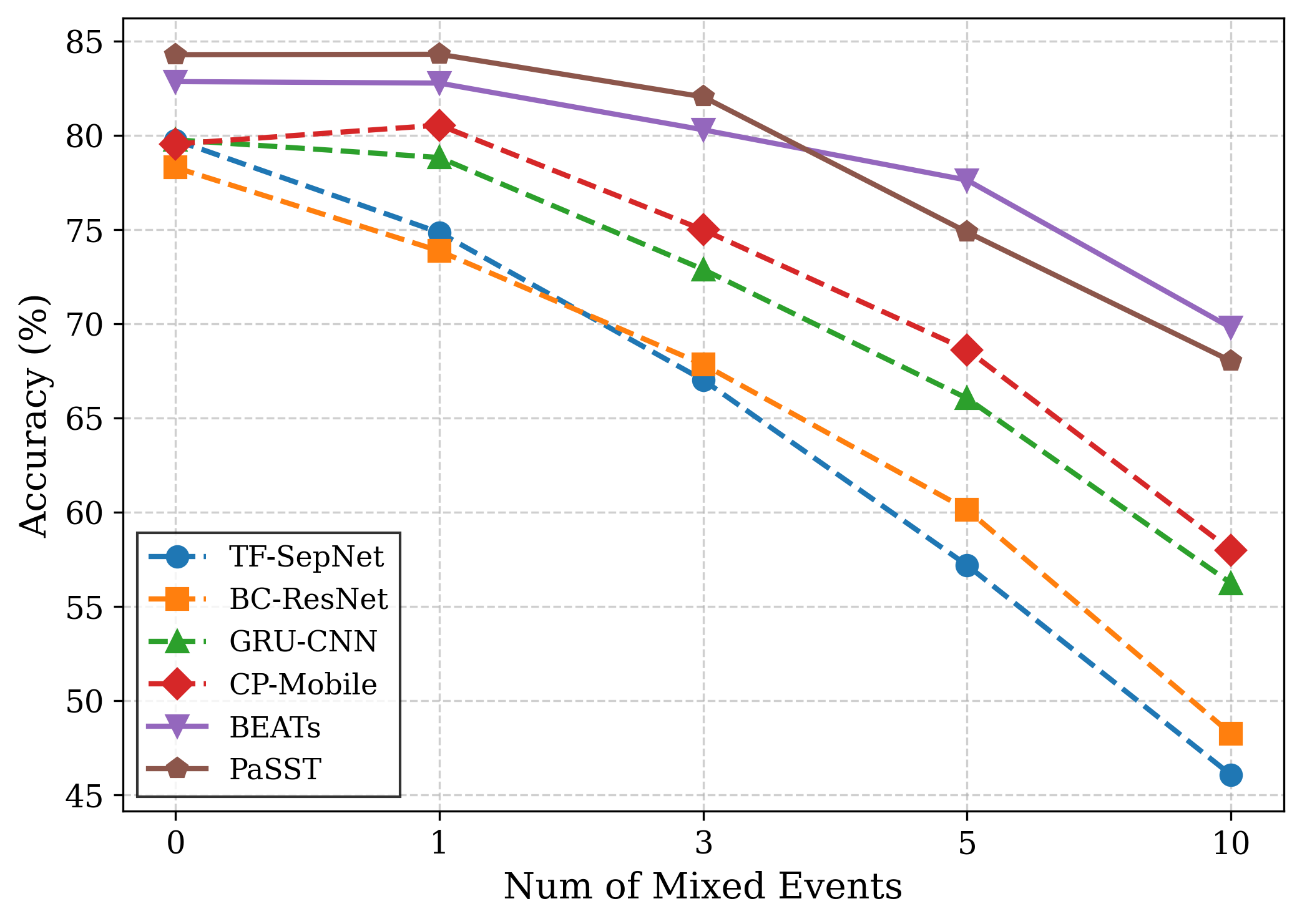}
    \caption{Classification accuracy (\%) under varying levels of event mixing. Dashed lines indicate lightweight CNNs, and solid lines denote externally pre-trained Transformers.}
    \label{fig:event_num}
\end{figure}

\noindent
\textbf{Impact of Mixed Event Number.}
Figure \ref{fig:event_num} illustrates the impact of increasing number of mixed events per clip (ranging from 0 to 10) on the classification performance. Under clean background conditions or low event polyphony (0 to 1 event), all models demonstrate strong performance, with PaSST and BEATs establishing the upper bound. However, as the acoustic scene becomes increasingly cluttered with diverse foreground events, a nearly monotonic degradation in accuracy becomes evident across all baselines. The lightweight models, particularly TF-SepNet and BC-ResNet, are the most susceptible to this escalating polyphony, suffering a significant accuracy drop from approximately 78-80\% down to below 50\% at the extreme condition of 10 mixed events. This dramatic decline highlights that dense acoustic complexity of sound events alone disrupts the representations learned by ASC models. Conversely, the externally pre-trained models exhibit markedly higher resilience. BEATs and PaSST manage to sustain accuracies near 68\% to 70\% even at the maximum complexity of 10 overlapping events. These findings underscore that while highly polyphonic foreground events uniformly degrade acoustic scene recognition, large-scale pre-training offer a noticeably strong defense against severe acoustic clutter.

\begin{figure}
    \centering
    \includegraphics[width=\linewidth]{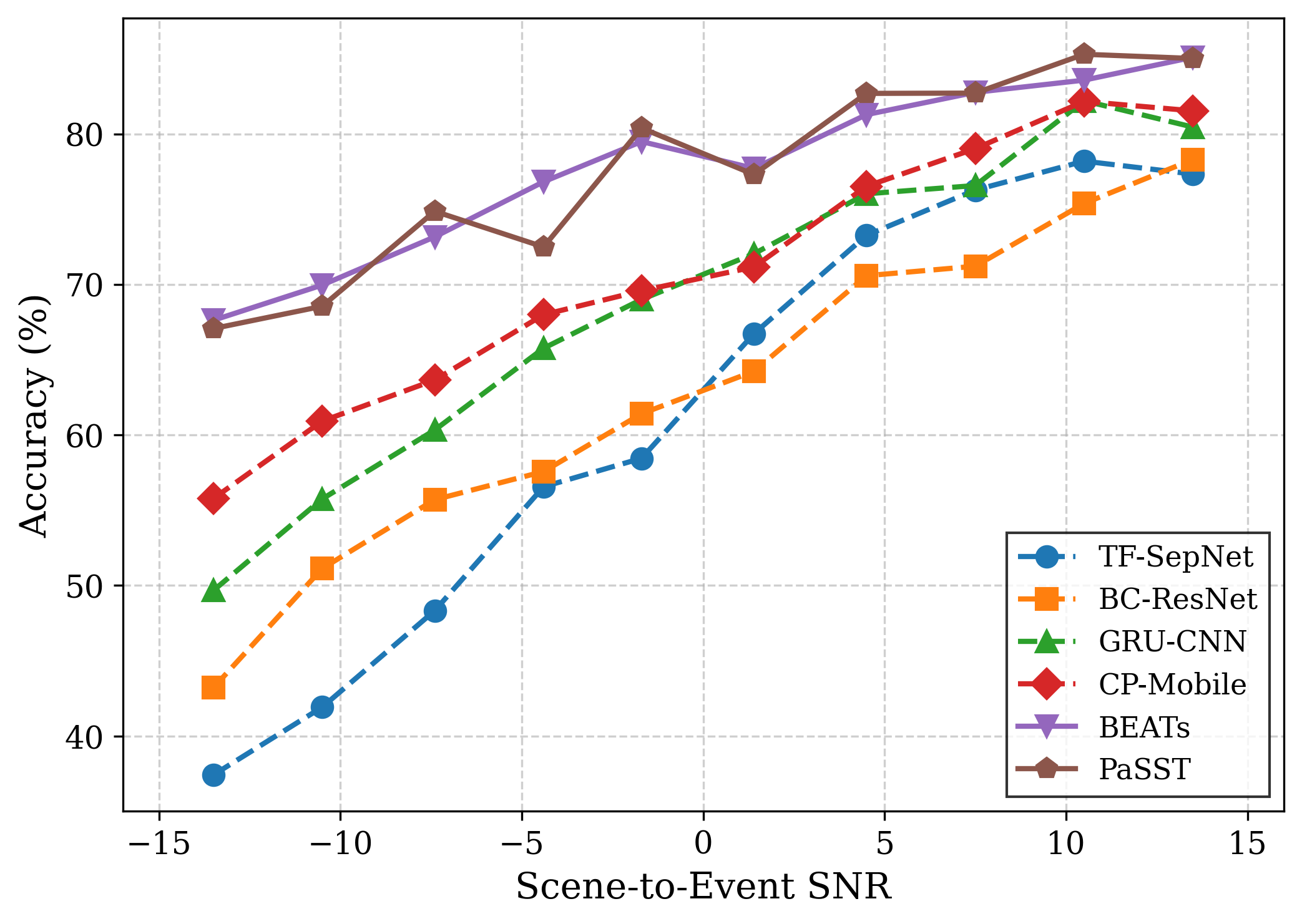}
    \caption{Classification accuracy (\%) under varying scene-to-event Signal-to-Noise Ratio (SNR) conditions.}
    \label{fig:mix_snr}
\end{figure}

\noindent
\textbf{Impact of Applied SNR.}
Figure \ref{fig:mix_snr} illustrates the influence of the scene-to-event Signal-to-Noise Ratio (SNR) on the classification accuracy of the evaluated systems. As the SNR decreases—meaning the injected foreground events become progressively louder and more dominant relative to the background scene—all models experience a notable degradation in performance. At high SNR levels, where the foreground events are relatively quiet, the models perform at peak, closely approaching their baseline capabilities on clean acoustic scenes. However, under the extreme negative SNR conditions, a stark divergence in model robustness emerges. The lightweight CNNs suffer a severe performance collapse, with TF-SepNet and BC-ResNet plummeting to 37.42\% and 43.21\% accuracy, respectively. In contrast, the large-scale Transformers demonstrate substantial resilience against the aggressive interference, retaining approximately 67\% accuracy even when at the worst condition. These findings reinforce the conclusion that while dominant foreground noises significantly disrupt standard acoustic representations, large-scale pre-training offers a effective defense against severe event interference.

\section{Conclusion}
In this paper, we introduced the Event-Shifted Acoustic Scene (ESAS) dataset, a new benchmark designed to evaluate the robustness of Acoustic Scene Classification (ASC) systems against unexpected foreground sounds. By evaluating various state-of-the-art models, we exposed a critical vulnerability in current ASC methods. Experimental results reveal two failure modes: models suffer a severe collapse in accuracy when encountering unknown events, and also degrade due to event crowdness in acoustic scenes. Furthermore, we observed that lightweight CNN architectures are more vulnerable to sound event interference compared to large-scale, pre-trained Transformer models. Ultimately, the ESAS dataset highlights the need to move beyond clean-audio training paradigms. We hope this benchmark will encourage future research toward fundamentally event-robust ASC models, particularly lightweight architectures capable of handling unpredictable real-world noise.

\section{Acknowledgments}
This work was supported by the Jiangsu Provincial Major Science and Technology Project (BG2024027).

\section{Generative AI Use Disclosure}
As detailed in the methodology section, OpenAI's GPT-4 was used to guide the semantic scene-event grouping during the construction of the ESAS dataset.

\bibliographystyle{IEEEtran}
\bibliography{mybib}

@article{barchiesi2015acoustic,
  title={Acoustic scene classification: Classifying environments from the sounds they produce},
  author={Barchiesi, Daniele and Giannoulis, Dimitrios and Stowell, Dan and Plumbley, Mark D},
  journal={IEEE Signal Processing Magazine},
  volume={32},
  number={3},
  pages={16--34},
  year={2015},
}

@inproceedings{Mesaros2018,
    author = "Mesaros, Annamaria and Heittola, Toni and Virtanen, Tuomas",
    title = "A multi-device dataset for urban acoustic scene classification",
    booktitle = "Proceedings of the Detection and Classification of Acoustic Scenes and Events 2018 Workshop (DCASE2018)",
    year = "2018",
    month = "November",
    pages = "9--13",
}

@inproceedings{Heittola2020,
    author = "Heittola, Toni and Mesaros, Annamaria and Virtanen, Tuomas",
    title = "Acoustic scene classification in {DCASE} 2020 Challenge: Generalization across devices and low complexity solutions",
    booktitle = "Proceedings of the Detection and Classification of Acoustic Scenes and Events 2020 Workshop (DCASE2020)",
    year = "2020",
    pages = "56--60",
}

@inproceedings{koutini22_interspeech,
  title     = {Efficient Training of Audio Transformers with Patchout},
  author    = {Khaled Koutini and Jan Schlüter and Hamid Eghbal-zadeh and Gerhard Widmer},
  year      = {2022},
  booktitle={Proceedings of the Conference of the International Speech Communication Association (INTERSPEECH)},
  organization = {ISCA},
  pages     = {2753--2757},
  doi       = {10.21437/Interspeech.2022-227},
  issn      = {2958-1796},
}

@inproceedings{Cai2024workshop,
    author = "Cai, Yiqiang and Li, Shengchen and Shao, Xi",
    title = "Leveraging Self-Supervised Audio Representations for Data-Efficient Acoustic Scene Classification",
    booktitle = "Proceedings of the Detection and Classification of Acoustic Scenes and Events 2024 Workshop (DCASE2024)",
    month = "October",
    year = "2024",
    pages = "21--25",
}

@inproceedings{schmid2022knowledge,
  title={Knowledge Distillation from Transformers for Low-Complexity Acoustic Scene Classification},
  author={Schmid, Florian and Masoudian, Shahed and Koutini, Khaled and Widmer, Gerhard},
  booktitle={Proceedings of the Detection and Classification of Acoustic Scenes and Events 2022 Workshop (DCASE2022)},
  year={2022}
}

@article{hou2023cooperative,
  title={Cooperative scene-event modelling for acoustic scene classification},
  author={Hou, Yuanbo and Kang, Bo and Mitchell, Andrew and Wang, Wenwu and Kang, Jian and Botteldooren, Dick},
  journal={IEEE/ACM transactions on audio, speech, and language processing},
  volume={32},
  pages={68--82},
  year={2023},
  publisher={IEEE}
}

@inproceedings{cai2024tf,
  title={{TF-SepNet}: An Efficient {1D} Kernel Design in {CNNs} for Low-Complexity Acoustic Scene Classification},
  author={Cai, Yiqiang and Zhang, Peihong and Li, Shengchen},
  booktitle={2024 IEEE International Conference on Acoustics, Speech and Signal Processing (ICASSP)},
  pages={821--825},
  year={2024},
  organization={IEEE}
}

@article{koutini2021receptive,
  title={Receptive field regularization techniques for audio classification and tagging with deep convolutional neural networks},
  author={Koutini, Khaled and Eghbal-zadeh, Hamid and Widmer, Gerhard},
  journal={IEEE/ACM Transactions on Audio, Speech, and Language Processing},
  volume={29},
  pages={1987--2000},
  year={2021},
  publisher={IEEE}
}

@inproceedings{Schmid2023,
    author = "Schmid, Florian and Morocutti, Tobias and Masoudian, Shahed and Koutini, Khaled and Widmer, Gerhard",
    title = "Distilling the Knowledge of Transformers and {CNNs} with {CP}-Mobile",
    booktitle = "Proceedings of the 8th Detection and Classification of Acoustic Scenes and Events 2023 Workshop (DCASE2023)",
    address = "Tampere, Finland",
    month = "September",
    year = "2023",
    pages = "161--165",
}

@InProceedings{pmlr-v202-chen23ag,
  title = 	 {{BEAT}s: Audio Pre-Training with Acoustic Tokenizers},
  author =       {Chen, Sanyuan and Wu, Yu and Wang, Chengyi and Liu, Shujie and Tompkins, Daniel and Chen, Zhuo and Che, Wanxiang and Yu, Xiangzhan and Wei, Furu},
  booktitle = 	 {Proceedings of the 40th International Conference on Machine Learning (ICML)},
  pages = 	 {5178--5193},
  year = 	 {2023},
  organization =    {PMLR},
}

@article{devalraju2022multiview,
  title={Multiview embeddings for soundscape classification},
  author={Devalraju, Dhanunjaya Varma and Rajan, Padmanabhan},
  journal={IEEE/ACM Transactions on Audio, Speech, and Language Processing},
  volume={30},
  pages={1197--1206},
  year={2022},
  publisher={IEEE}
}

@inproceedings{kim21l_interspeech,
  author={Byeonggeun Kim and Simyung Chang and Jinkyu Lee and Dooyong Sung},
  title={{Broadcasted Residual Learning for Efficient Keyword Spotting}},
  year=2021,
  booktitle={Proceedings of the Conference of the International Speech Communication Association (INTERSPEECH)},
  organization = {ISCA},
  pages={4538--4542},
}

@inproceedings{bear19_interspeech,
  title     = {Towards Joint Sound Scene and Polyphonic Sound Event Recognition},
  author    = {Helen L. Bear and Inês Nolasco and Emmanouil Benetos},
  year      = {2019},
  booktitle={Proceedings of the Conference of the International Speech Communication Association (INTERSPEECH)},
  organization = {ISCA},
  pages     = {4594--4598},
}

@book{virtanen2018computational,
  title={Computational analysis of sound scenes and events},
  author={Virtanen, Tuomas and Plumbley, Mark D and Ellis, Dan},
  volume={9},
  year={2018},
  publisher={Springer}
}

@inproceedings{jeong2022cochlscene,
  title={Cochlscene: Acquisition of acoustic scene data using crowdsourcing},
  author={Jeong, Il-Young and Park, Jeongsoo},
  booktitle={2022 Asia-Pacific Signal and Information Processing Association Annual Summit and Conference (APSIPA ASC)},
  pages={17--21},
  year={2022},
  organization={IEEE}
}

@article{bai2024description,
  title={Description on ieee icme 2024 grand challenge: Semi-supervised acoustic scene classification under domain shift},
  author={Bai, Jisheng and Wang, Mou and Liu, Haohe and Yin, Han and Jia, Yafei and Huang, Siwei and Du, Yutong and Zhang, Dongzhe and Shi, Dongyuan and Gan, Woon-Seng and others},
  journal={arXiv preprint arXiv:2402.02694},
  year={2024}
}

@inproceedings{bai2025apsipa,
  title={The APSIPA ASC 2025 Grand Challenge on City and Time-Aware Semi-Supervised Acoustic Scene Classification: Summary and Results},
  author={Bai, Jisheng and Wang, Mou and Liu, Haohe and Xiang, Bin and Liu, Ying and Chen, Jianfeng and Shi, Dongyuna and Plumbley, Mark D and Rahardja, Susanto and Gan, Woon-Seng},
  booktitle={2025 Asia Pacific Signal and Information Processing Association Annual Summit and Conference (APSIPA ASC)},
  pages={193--197},
  year={2025},
  organization={IEEE}
}

@article{tonami2021joint,
  title={Joint analysis of sound events and acoustic scenes using multitask learning},
  author={Tonami, Noriyuki and Imoto, Keisuke and Yamanishi, Ryosuke and Yamashita, Yoichi},
  journal={IEICE TRANSACTIONS on Information and Systems},
  volume={104},
  number={2},
  pages={294--301},
  year={2021},
  publisher={The Institute of Electronics, Information and Communication Engineers}
}

@article{fonseca2021fsd50k,
  title={Fsd50k: an open dataset of human-labeled sound events},
  author={Fonseca, Eduardo and Favory, Xavier and Pons, Jordi and Font, Frederic and Serra, Xavier},
  journal={IEEE/ACM Transactions on Audio, Speech, and Language Processing},
  volume={30},
  pages={829--852},
  year={2021},
  publisher={IEEE}
}

@article{zhang2025ddsc,
  title={DDSC: Dynamic Dual-Signal Curriculum for Data-Efficient Acoustic Scene Classification under Domain Shift},
  author={Zhang, Peihong and Liu, Yuxuan and Sang, Rui and Li, Zhixin and Cai, Yiqiang and Tan, Yizhou and Li, Shengchen},
  journal={arXiv preprint arXiv:2510.17345},
  year={2025}
}

@article{tan2024acoustic,
  title={Acoustic scene classification across cities and devices via feature disentanglement},
  author={Tan, Yizhou and Ai, Haojun and Li, Shengchen and Plumbley, Mark D},
  journal={IEEE/ACM Transactions on Audio, Speech, and Language Processing},
  volume={32},
  pages={1286--1297},
  year={2024},
  publisher={IEEE}
}

@techreport{Tan2025,
    Author = "Tan, Ee-Leng and Yeow, Jun Wei and Peksi, Santi and Li, Haowen and Yang, Ziyi and Gan, Woon-Seng",
    title = "SNTL-Ntu Dcase25 Submission: Acoustic Scene Classification Using {CNN}-{GRU} Model Without Knowledge Distillation",
    institution = "DCASE2025 Challenge",
    year = "2025",
    month = "May",
}

\end{document}